\documentclass[preprint2,twocolumn,times,tighten]{aastex631}

\usepackage{graphicx}
\usepackage{amsmath}
\usepackage{amssymb}
\let\tablenum\relax
\usepackage{siunitx}

\newcommand{\nh}{n_{\rm H}}

\newcommand{\pc}{\mathrm{pc}}

\begin{document}

\title{Intermittency in the Multiphase ISM --- II. Non-Gaussian Dust Polarization and Foreground Variance}

\email{yuehu@caltech.edu}

\author[0000-0002-8455-0805]{Yue Hu}
\affiliation{Cahill Center for Astronomy and Astrophysics, California Institute of Technology, Pasadena, CA 91125, USA}

\begin{abstract}
Interstellar turbulence produces non-Gaussian polarized Galactic dust emission whose higher-order statistics depend on thermal phase and turbulent forcing. We synthesize optically thin dust polarization from three thermally multiphase diffuse interstellar medium (ISM) simulations with compressive, mixed, and solenoidal forcing. We characterize the resulting non-Gaussian structure in the warm, unstable, and cold phases using $E$- and $B$-mode spectra, one-point distributions, scale-dependent increments, and Minkowski functionals. Both $E$ and $B$ modes are intermittent. Warm-phase emission remains closest to Gaussian, whereas cold gas shows flatter intermediate-scale spectra, broader non-Gaussian tails, and the strongest departures from Gaussian morphology. Compressive forcing generally produces the strongest projected intermittency, particularly in the cold phase and total emission, while mixed and solenoidal forcing show weaker signatures. These higher-order structures also enhance spatial fluctuations in polarized dust power: Gaussian fields matched to the same two-point covariance underestimate the region-to-region scatter by factors of up to $\sim2.6$. Our results show that thermal phase and turbulent forcing regulate not only the mean polarized dust spectrum but also the field-to-field distribution of foreground power, with implications for modeling Galactic foreground in CMB polarization measurements.
\end{abstract}


\section{Introduction}\label{sec:introduction}
Primordial gravitational waves are expected to leave a distinctive large-scale curl pattern in the polarization of the cosmic microwave background (CMB) \citep{ZaldarriagaSeljak1997,KamionkowskiKosowskyStebbins1997, KamionkowskiKovetz2016}. Detecting this signal, usually characterized by the tensor-to-scalar ratio $r$, would provide a direct window into the physics of the early Universe. The difficulty is that the same sky contains polarized emission from Galactic dust, which remains one of the dominant foregrounds at the frequencies and angular scales targeted by current CMB experiments \citep{Planck2018XI2020,BICEPKeck2021}.

Dust polarization is not a smooth or statistically simple foreground. Its Stokes parameters arise from the line-of-sight combination of gas density, magnetic-field strength and orientation, and the alignment of nonspherical dust grains with the magnetic field \citep{DraineFraisse2009,HoangLazarian2008,Andersson2015, PlanckIntXIX2015,Vansyngel2017,2025ApJ...994..115H}. These quantities are coupled by interstellar turbulence, whose density and magnetic fluctuations develop scale-dependent structure and anisotropy \citep{GoldreichSridhar1995,ChoLazarian2002,ChoLazarian2003, Federrath2010,Burkhart2018,2019ApJ...878..157X,2021ApJ...911...37H,2025ApJ...986...62H}. Observationally, polarized dust emission is organized into coherent structures whose orientations are correlated with the Galactic magnetic field \citep{PlanckIntXXXII2016,PlanckIntXXXVIII2016,Soler2013,2026ApJ..1000...37P} and turbulence \citep{2023MNRAS.519.3736H,2023MNRAS.524.2994H}. Both observations and synthetic polarization maps therefore contain non-Gaussian, multiscale structure beyond that described by their power spectra \citep{RegaldoSaintBlancard2020,Delouis2022}.

Most descriptions of polarized dust nevertheless begin with its two-point statistics. At high Galactic latitude, the dust $E$- and $B$-mode spectra are approximately power laws, and their amplitudes and slopes provide a useful description of the mean foreground \citep{PlanckIntXXX2016,PlanckIntXXXVIII2016,Planck2018XI2020}. Analytic models and MHD simulations have connected these spectra and their $E/B$ asymmetry to turbulent anisotropy, compressibility, magnetic-field geometry, and density--field alignment \citep{Caldwell2017,Kandel2018,Kritsuk2018,Vansyngel2017, 2026ApJ..1007..138H}. A power spectrum, however, provides a complete statistical description only for a Gaussian field. For an intermittent
foreground, two skies can have the same mean $BB$ spectrum while differing substantially in the strength, morphology, and spatial concentration of their fluctuations.

This distinction matters because CMB experiments observe finite regions of the sky rather than an ensemble average. Rare structures and coherent filaments can concentrate dust power into particular regions, producing field-to-field fluctuations that are not determined by the mean power spectrum alone \citep{RegaldoSaintBlancard2020,AbrilCabezas2024}. Intermittency provides a natural physical origin for this behavior. In turbulent flows, increasingly non-Gaussian fluctuations toward small scales reflect the concentration of strong gradients into a small fraction of the volume \citep{SheLeveque1994,PolitanoPouquet1995}. In the multiphase ISM, thermal instability and turbulent compression further redistribute these fluctuations among the warm, unstable, and cold gas \citep{McKeeOstriker1977,Wolfire2003,Hu2026PaperI}. Thus, reproducing the mean dust power spectrum does not necessarily reproduce the probability of encountering an unusually contaminated observing field.

The physical origin of this intermittency was explored in our companion study, Paper~I \citep{Hu2026PaperI}. Using thermally multiphase MHD simulations, we showed that density, velocity, and magnetic-field fluctuations are all non-Gaussian and scale-dependent, but their responses to thermal phase and turbulent forcing differ. Magnetic intermittency shows the clearest forcing dependence, with the strongest phase-normalized magnetic intermittency occurring in the unstable neutral medium (UNM). Compressive driving also increases the cold-neutral-medium (CNM) mass fraction and redistributes density and magnetic power preferentially into cold gas. Extreme magnetic increments occur preferentially where magnetic curl and magnetic tension are simultaneously enhanced. These results raise a natural next question: how much of this three-dimensional intermittent structure survives density weighting and line-of-sight projection into observable dust polarization?

In this work, we address this question by synthesizing polarized dust emission from the same compressively, mixed, and solenoidally driven simulations and separating the contributions from the warm neutral medium (WNM), UNM, and CNM. We examine how thermal phase and turbulent forcing shape the non-Gaussian statistics of dust polarization $E$ and $B$ modes, and how those statistics translate into spatial fluctuations of foreground power. Our central comparison is between the simulated dust maps and Gaussian fields constructed to have the same two-point covariance. This isolates the additional variance carried by higher-order structure while keeping the mean dust power fixed, complementing previous statistical studies of non-Gaussian polarized dust \citep{RegaldoSaintBlancard2020,Delouis2022,AbrilCabezas2024}.

We characterize this structure using one-point distributions, scale-dependent increments, and Minkowski functionals, and quantify its impact on the field-to-field variation of dust band powers. The latter is also expressed as an equivalent contamination of $r$ to connect the result with CMB $B$-mode measurements. Instrumental noise, lensing $B$ modes, beam effects, frequency decorrelation, and component-separation uncertainties are beyond the scope of this study.

We use the same $1024^3$ compressive, mixed, and solenoidal forcing suite as Paper~I and average the statistics over three statistically independent snapshots. A uniform-alignment polarization prescription allows us to isolate the effects of turbulent intermittency and thermal phase under controlled conditions. Section~\ref{sec:methods} describes the simulations and observables, Section~\ref{sec:results} presents the forcing and phase dependence, Section~\ref{sec:discussion} discusses the implications for ISM physics and CMB foregrounds, and Section~\ref{sec:conclusions} summarizes our main conclusions.

\section{Methods}\label{sec:methods}

\subsection{Multiphase MHD simulations}\label{sec:simulations}

We analyze the three simulations introduced in Paper~I \citep{Hu2026PaperI}. Each evolves ideal compressible MHD with AthenaK \citep{2024arXiv240916053S,2025ApJ...986...62H} in a periodic cube of side \(L=100\,\pc\) on a \(1024^3\) uniform grid. The initial mean hydrogen density is \(\nh=1\,\mathrm{cm}^{-3}\), and the initially uniform magnetic field has strength approximately 3~\SI{}{\micro G}, consistent with Zeeman observations \citep{2012ARA&A..50...29C}. The velocity dispersion is $\sigma_v \approx 10$~km~s$^{-1}$, based on observational constraints derived from Larson's law and young stellar environments \citep{1981MNRAS.194..809L,2022ApJ...934....7H}. Atomic heating and cooling produce a thermally multiphase neutral medium following \citet{2002ApJ...564L..97K}. Turbulence is driven on scales of 50--100 pc by an acceleration field whose solenoidal fraction is \(f_{\rm sol}=0\), 0.5, or 1 for compressive, mixed, and solenoidal forcing, respectively. Apart from this projection of the driving field, the numerical setup is identical among the runs.

We use three snapshots from the statistically developed state of every run, separated by 5 Myr, corresponding to one eddy-turnover time. Unless noted otherwise, curves and symbols show the mean over the three snapshots, and shaded regions or error bars show their sample standard deviation.

Following the phase convention of Paper~I \citep{Hu2026PaperI}, we divide the gas by temperature into the CNM ($T<200$~K), UNM ($200\leq T<5000$~K), and WNM ($T\geq5000$~K). Temperature masks are applied cell by cell before the line-of-sight integration, so each phase-restricted map retains the density contrast, magnetic-field orientation, and filling factor produced by the simulation.
The total projection gives the physical superposition of all phases, whereas the phase-restricted maps identify which gas component is responsible for particular statistical features.

\subsection{Synthetic dust polarization}\label{sec:stokes}
We construct optically thin thermal-dust polarization maps at 353 GHz. The dust temperature is fixed at $T_{\rm d}=19.6$ K, and the opacity per hydrogen nucleus is $8\times10^{-27}\,\mathrm{cm^2\,H^{-1}}$, consistent with the single-temperature modified-blackbody prescription commonly adopted in \textit{Planck} foreground analyses \citep{PlanckIntXIX2015,PlanckIntXXX2016,Planck2018XI2020}. For fixed dust temperature, opacity, and dust-to-gas ratio, the optically thin emissivity scales as
\begin{equation}
j_\nu \propto n_{\rm H}\,\sigma_{\nu,{\rm H}}\,B_\nu(T_{\rm d})
\propto n_{\rm H},
\end{equation}
where $n_{\rm H}$ is the hydrogen nuclei number density and $\sigma_{\nu,{\rm H}}$ is the opacity per hydrogen nucleus. We further assume spatially uniform grain alignment and a constant intrinsic polarization fraction $p_0=0.3$ \citep{PlanckIntXIX2015,Vansyngel2017}. These assumptions isolate the polarization structure generated by the simulated gas density and magnetic field.

For a line of sight along the $z$-direction, the plane-of-sky magnetic-field components are $B_x$ and $B_y$. The optically thin Stokes parameters can then be written, up to the common factor $\sigma_{\nu,{\rm H}}B_\nu(T_{\rm d})$, as \citep{PlanckIntXIX2015,Vansyngel2017,2024ApJ...965..183H}
\begin{equation}
\begin{aligned}
I &\propto \int ds\,n_{\rm H}
\left[
1-p_0\left(
\frac{B_x^2+B_y^2}{B^2}-\frac{2}{3}
\right)
\right],\\
Q &\propto p_0\int ds\,n_{\rm H}
\frac{B_y^2-B_x^2}{B^2},\\
U &\propto -2p_0\int ds\,n_{\rm H}
\frac{B_xB_y}{B^2},
\end{aligned}
\label{eq:stokes}
\end{equation}
where $B^2=B_x^2+B_y^2+B_z^2$ and $s$ is the line-of-sight coordinate. We also define the polarized intensity $P=(Q^2+U^2)^{1/2}$, polarization fraction $p=P/I$, and polarization angle $\psi=\tfrac12\operatorname{atan2}(U,Q)$.

Eq.~\eqref{eq:stokes} shows that the polarized emission depends jointly on density weighting and magnetic-field geometry. Dense regions contribute
more strongly to the line-of-sight integral, while changes in projected field orientation can either reinforce or cancel the polarized signal. A dense structure with a coherent magnetic field can therefore dominate the polarization, whereas rapid changes in field orientation can lead to strong depolarization. The resulting maps thus reflect the coupled density and magnetic structure of the turbulent medium.


\begin{figure*}
\centering
\includegraphics[width=0.99\linewidth]{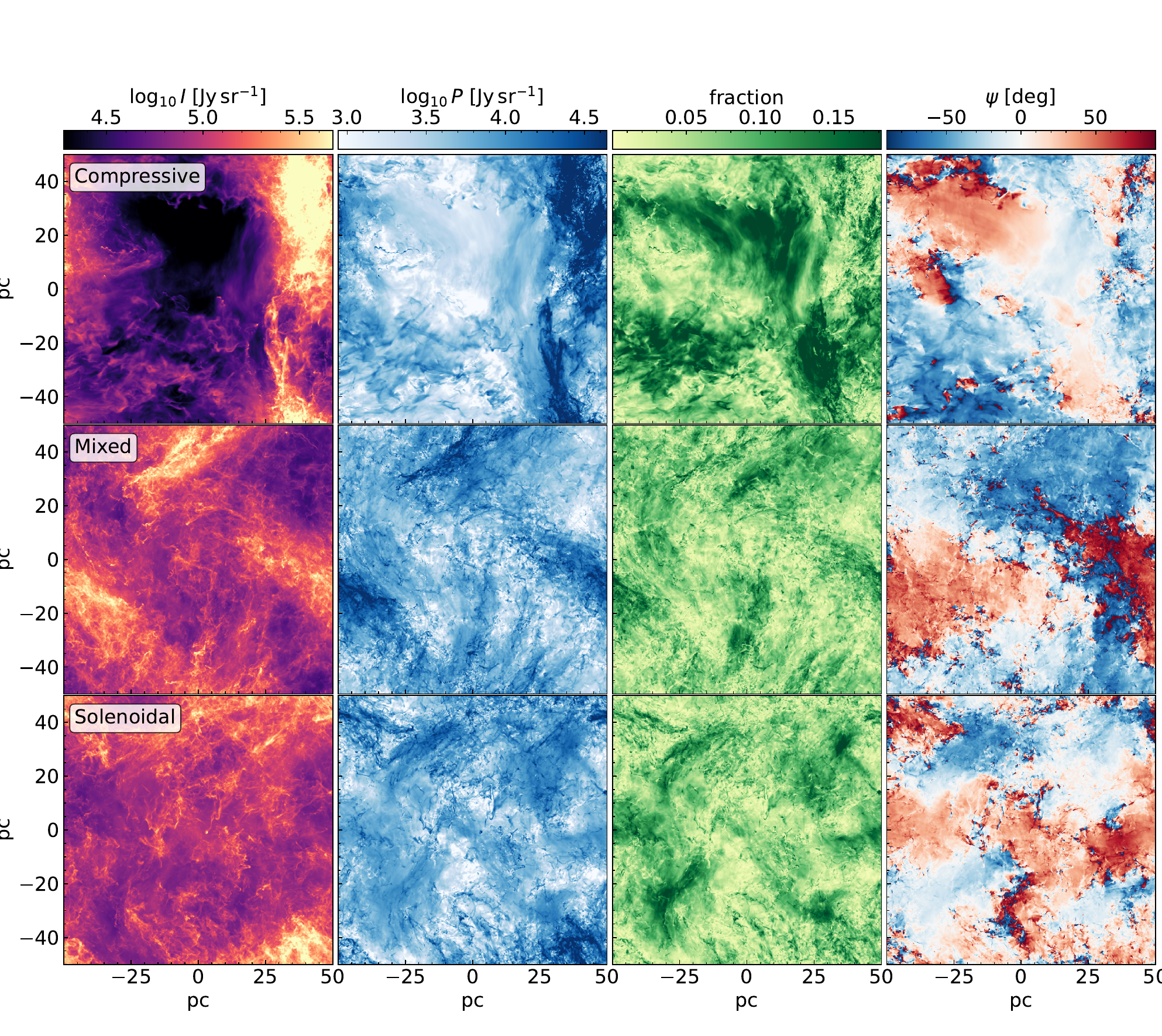}
\caption{Synthetic 353-GHz dust-polarization observables for one representative snapshot viewed along \(z\). Rows show compressive, mixed, and solenoidal forcing; columns show total intensity \(I\), polarized intensity \(P\), polarization fraction \(p\), and polarization angle \(\psi\). }
\label{fig:polmaps}
\end{figure*}

\subsection{Flat-sky \(E\) and \(B\) modes}
\label{sec:eb}
For each periodic polarization map, we Fourier transform the Stokes parameters $Q$ and $U$. If $\varphi_{\boldsymbol{k}}$ is the polar angle of the two-dimensional wavevector $\boldsymbol{k}$, the flat-sky decomposition is
\begin{equation}
\begin{aligned}
\widetilde E(\boldsymbol{k})
&=
\widetilde Q(\boldsymbol{k})\cos 2\varphi_{\boldsymbol{k}}
+
\widetilde U(\boldsymbol{k})\sin 2\varphi_{\boldsymbol{k}},\\
\widetilde B(\boldsymbol{k})
&=
-\widetilde Q(\boldsymbol{k})\sin 2\varphi_{\boldsymbol{k}}
+
\widetilde U(\boldsymbol{k})\cos 2\varphi_{\boldsymbol{k}}.
\end{aligned}
\label{eq:eb}
\end{equation}
This corresponds to the flat-sky limit of the standard parity decomposition \citep{ZaldarriagaSeljak1997}. We inverse transform $\widetilde E$ and $\widetilde B$ to obtain real-space $E$- and $B$-mode maps. For the subregion analysis, the isotropized band powers $C^{XX}(k)$, with $X\in\{E,B\}$, are computed by averaging $|\widetilde X|^2$ over annuli of physical wavenumber $k$, expressed in cycles per parsec.

Because the $E/B$ decomposition is non-local in real space, a localized polarization structure generally contributes to both modes according to its geometry and orientation. We therefore interpret the real-space $E$ and $B$
maps statistically rather than associating individual structures uniquely with either mode. Their one-point and increment statistics retain information on the amplitude, scale dependence, and spatial concentration of extreme polarization fluctuations.

\begin{figure*}
\centering
\includegraphics[width=0.95\linewidth]{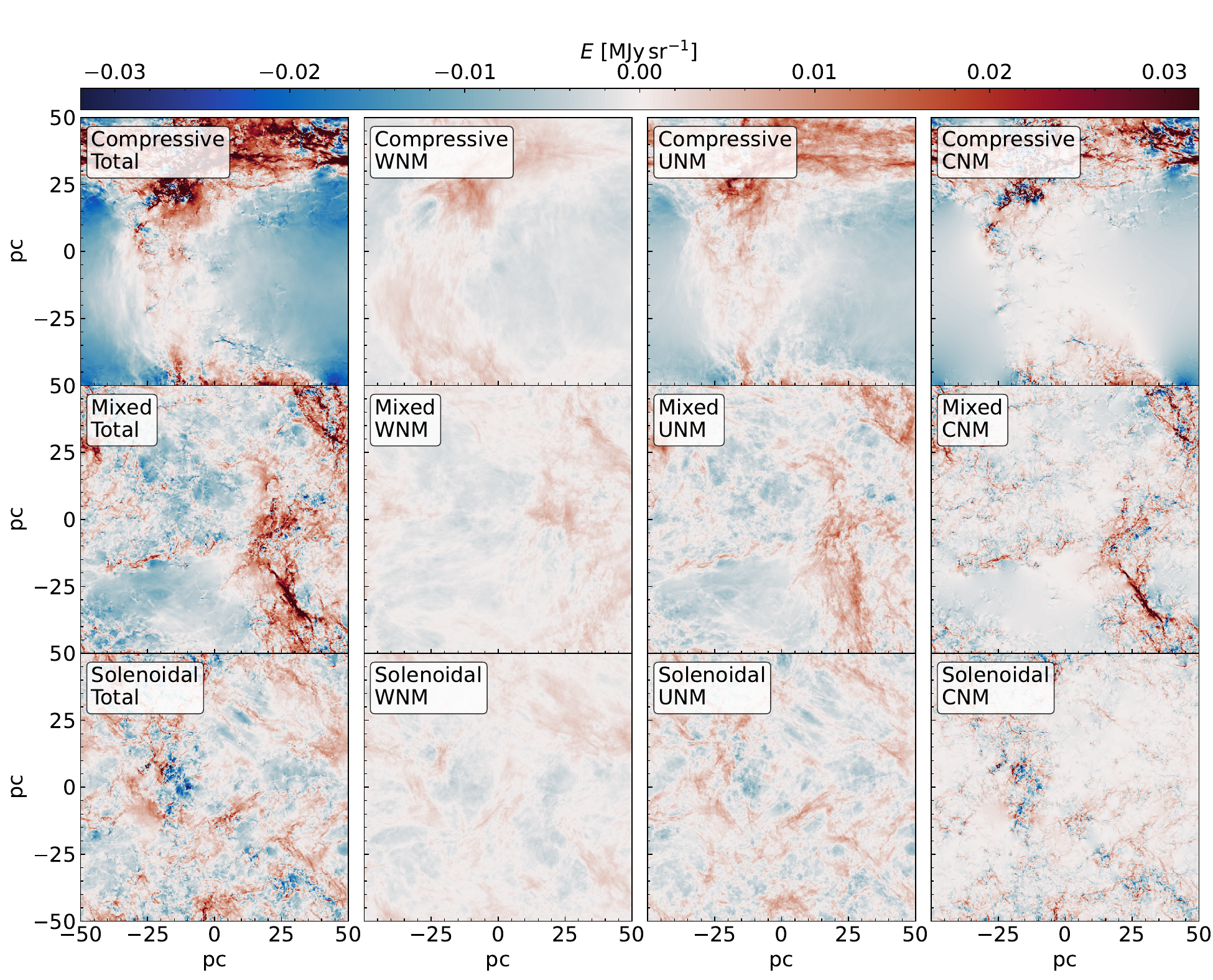}
\caption{Real-space dust \(E\)-mode maps at 353 GHz for the representative middle snapshot viewed along \(z\). Rows show compressive, mixed, and solenoidal forcing; columns show total, WNM, UNM, and CNM emission. All panels share the same symmetric color range.}
\label{fig:emaps}
\end{figure*}

\begin{figure*}
\centering
\includegraphics[width=0.95\linewidth]{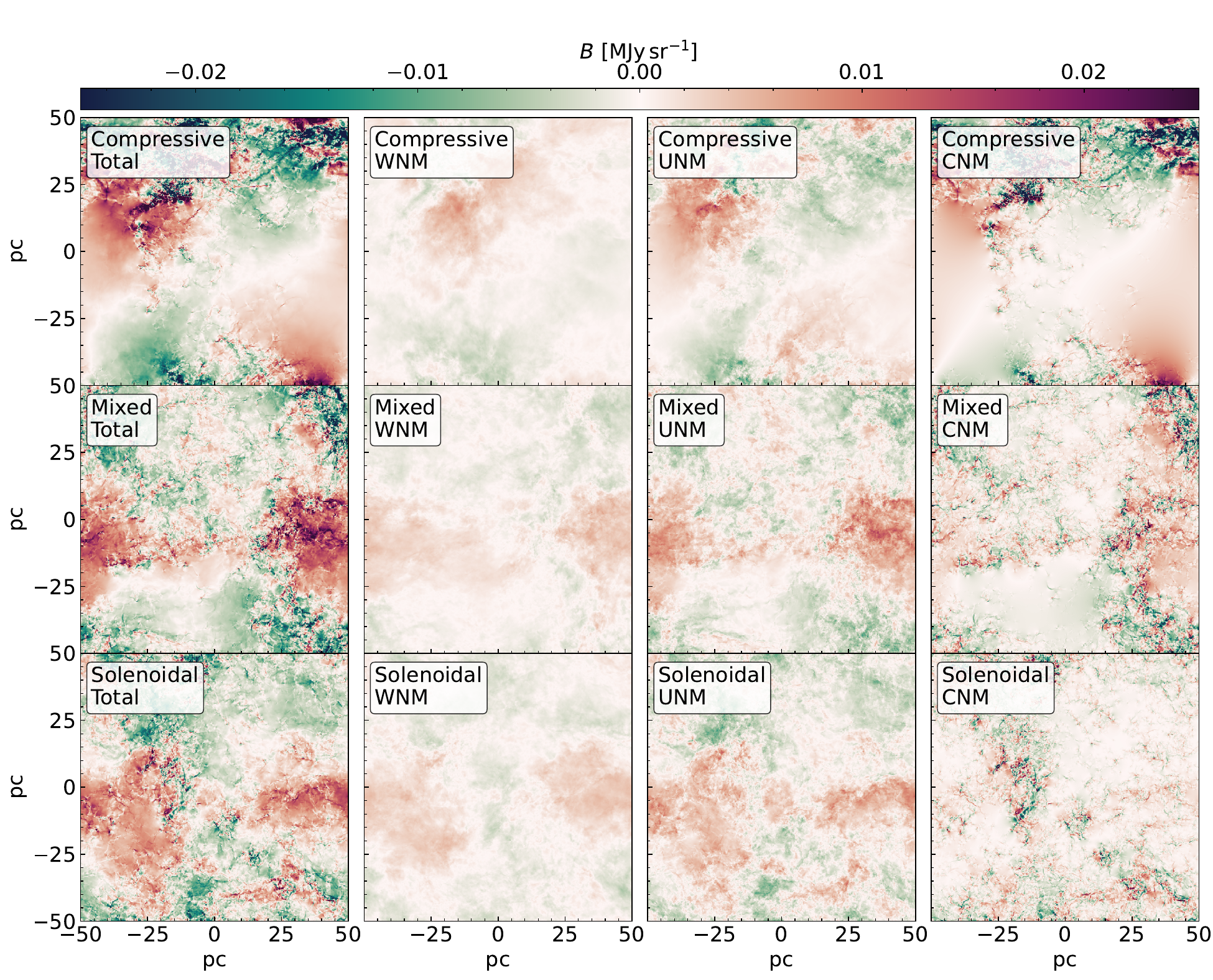}
\caption{Real-space dust \(B\)-mode maps for the snapshot and line of sight used in Fig.~\ref{fig:emaps}.}
\label{fig:bmaps}
\end{figure*}

\subsection{Intermittency statistics}
\label{sec:statistics}
We quantify the non-Gaussian statistics of the projected polarization maps using one-point PDFs, scale-dependent increments, and two-dimensional Minkowski functionals, following the framework of Paper~I \citep{Hu2026PaperI}. For standardized maps $X/\sigma_X$, the one-point PDF characterizes the occurrence of extreme amplitudes.

\textbf{Scale-dependent intermittency} is measured from map increments,
\begin{equation}
\delta_{\ell}X(\boldsymbol{x}) =
X(\boldsymbol{x}+\boldsymbol{\ell})-X(\boldsymbol{x}),
\label{eq:increments}
\end{equation}
normalized by their rms at each lag. We evaluate the increment PDFs at $\ell=1$, 5, 10, and 20 pc. The development of progressively heavier tails toward smaller $\ell$ indicates increasing intermittency toward small scales and provides the projected counterpart of the scale-dependent behavior found in Paper~I.

\textbf{Spatial morphology} is quantified using the three two-dimensional Minkowski functionals of the excursion set $X/\sigma_X>\nu$: the area fraction $V_0$, contour length $V_1$, and Euler characteristic $V_2$ \citep{SchmalzingGorski1998}. We compare each map with random-phase Gaussian realizations that preserve its Fourier amplitudes. Deviations in $V_i(\nu)$ therefore quantify higher-order spatial organization not contained in the power spectrum.

The three statistics probe distinct aspects of non-Gaussianity: the one-point PDFs measure the abundance of extreme fluctuations, the increment PDFs their scale dependence, and the Minkowski functionals their morphology and connectivity. Together, they characterize both the amplitude and spatial organization of intermittency.

The PDFs and Minkowski functionals are evaluated along the $z$ line of sight and averaged over three snapshots. Because the simulations are  super-Alfv\'enic, the turbulent magnetic field is weakly constrained by the mean-field direction on the scales considered here, and line-of-sight dependence is correspondingly weak. We therefore adopt a fixed projection for the forcing- and phase-dependent comparisons. The power-spectrum and ubregion analyses combine the three orthogonal lines of sight to increase statistical sampling.

\subsection{Gaussian control and spatial foreground variance}
\label{sec:null}
To isolate the contribution of non-Gaussian structure to foreground variance, we compare the MHD polarization maps with Gaussian realizations matched to their measured two-point covariance. The full $100$-pc $E/B$ maps are first constructed and then divided into sixteen non-overlapping $25$-pc subregions, so that the non-local $E/B$ decomposition is performed before introducing subregion boundaries. The same mean subtraction, apodization, and band-power estimator are applied to the MHD and Gaussian maps.

The Gaussian controls preserve the measured Fourier covariance while removing higher-order phase correlations. Differences in the subregion-to-subregion power scatter therefore quantify information not contained in the two-point
statistics. We characterize this excess variance by $\sigma_{\rm MHD}^{XX}/\sigma_{\rm Gaussian}^{XX}$, with $X\in\{E,B\}$,
where $\sigma_{\rm MHD}^{XX}/\sigma_{\rm Gaussian}^{XX}=1$ corresponds to the Gaussian expectation, while $\sigma_{\rm MHD}^{XX}/\sigma_{\rm Gaussian}^{XX}>1$ indicates enhanced spatial variance from non-Gaussian structure.

To express the same effect on a cosmological scale, we map physical wavenumber to angular multipole using $\ell=2\pi d k,$ with $d=286.48$ pc, so that each $25$-pc subregion subtends $5^\circ$. The mean dust amplitude is normalized at 353 GHz. For each subregion, the deviation of the dust $BB$ spectrum from the parent mean is projected onto an $r=1$ primordial tensor template over $60\leq\ell<360$. We denote the resulting equivalent contamination by $\delta r_{\rm eq}$ and compare its subregion scatter, $\sigma(\delta r_{\rm eq})$, between the MHD and Gaussian ensembles.

This construction isolates the additional field-to-field foreground variance produced by higher-order dust structure at fixed mean power and two-point covariance.

\section{Results}\label{sec:results}

\subsection{Forcing and phase reorganize the projected maps}\label{sec:mapresults}
Fig.~\ref{fig:polmaps} shows clear differences in the projected dust structure among the three forcing cases. Compressive forcing produces the strongest intensity contrast, with bright structures concentrated in a relatively small fraction of the map. The mixed and solenoidal cases contain more spatially distributed filamentary structures and weaker large-scale contrast. These differences are also visible in polarized intensity and provide a qualitative indication of the forcing dependence of the non-Gaussian statistics discussed below.

The polarized intensity $P$ broadly follows the main structures in $I$, but is additionally affected by magnetic-field orientation along the line of sight. The polarization fraction $p$ therefore varies according to the degree of field coherence and cancellation, while the polarization angle $\psi$ shows extended regions of similar orientation separated by relatively sharp changes. The compressive case shows larger coherent variations in both $p$ and $\psi$, whereas the mixed and solenoidal cases show more distributed small-scale variation. The projected polarization thus reflects the combined effects of density structure and magnetic-field geometry.

Figs.~\ref{fig:emaps} and \ref{fig:bmaps} show the corresponding real-space $E$- and $B$-mode maps. In the compressive case, the strongest $E$ and $B$ fluctuations are concentrated in a small number of regions associated with the strongest structures in the polarization maps. Mixed forcing produces more extended filamentary fluctuations, while the solenoidal case distributes the fluctuations more broadly across the map. Because the $E/B$ transform is non-local, the positive and negative features reflect the spatial variation of the polarization field rather than separate emission components.

The phase-separated maps show a change from relatively smooth WNM emission to increasingly sharp structure in the colder phases. The UNM contains stronger and more coherent fluctuations than the WNM, while the CNM contains the narrowest filaments and many of the largest amplitudes seen in the total maps. This contribution is particularly clear in the compressive case, where the strongest structures in the total $E$ and $B$ maps are also prominent in the CNM maps.

This phase dependence differs from the intrinsic intermittency measured in Paper~I \citep{Hu2026PaperI}. There, increments were normalized within each thermal phase, and the UNM showed the strongest magnetic intermittency relative to its own variance. Here, dust polarization is density-weighted before projection, so dense CNM structures can make the largest contribution to the absolute polarized fluctuations. The two results therefore measure different aspects of the same multiphase turbulence: the UNM has stronger normalized magnetic intermittency, whereas the CNM contributes more strongly to the largest emission-weighted structures in the dust maps.

\begin{figure*}
\centering
\includegraphics[width=0.95\linewidth]{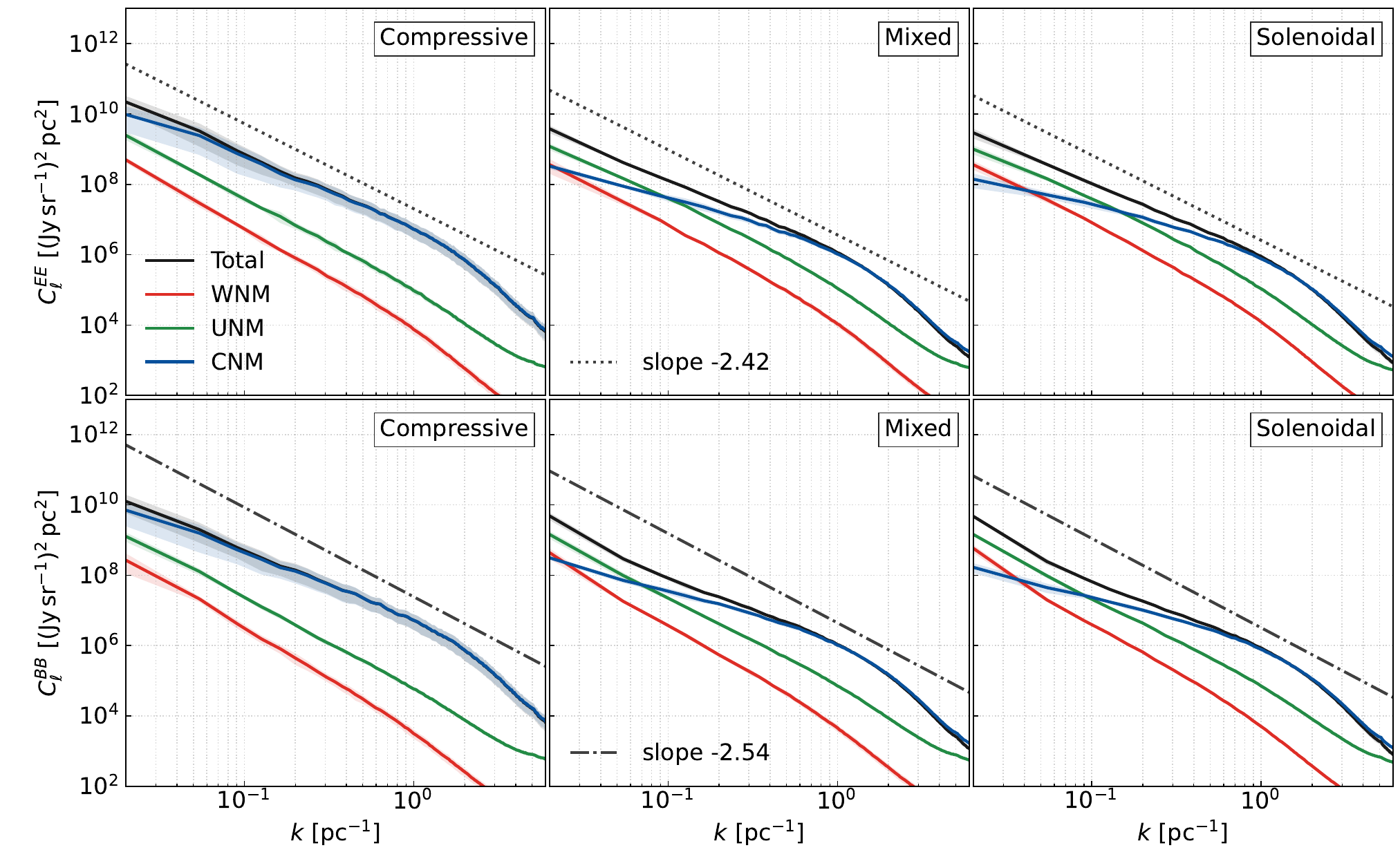}
\caption{Dust-polarization power spectra in physical wavenumber. Columns show compressive, mixed, and solenoidal forcing; rows show \(EE\) and \(BB\). Black, red, green, and blue denote total, WNM, UNM, and CNM emission. The gray reference lines have slopes \(-2.42\) for \(EE\) and \(-2.54\) for \(BB\), corresponding to the \textit{Planck} measurements.}
\label{fig:spectra}
\end{figure*}

\subsection{Power spectra and the scale distribution of structure}
\label{sec:spectraresults}
Fig.~\ref{fig:spectra} shows the $EE$ and $BB$ power spectra for the total emission and the three thermal phases. The spectra are not described by a single power law over the full resolved range: their slopes vary with wavenumber, and all components steepen toward the numerical dissipation range \citep{2026ApJ..1007..138H}. We therefore use the spectra mainly to compare the relative scale dependence among phases and forcing cases. The dotted and dash-dotted lines show the \textit{Planck} reference slopes, $-2.42$ for $EE$ and $-2.54$ for
$BB$ \citep{PlanckIntXXX2016}, for visual comparison only.

A clear phase ordering is present in both $EE$ and $BB$. The WNM spectra are the steepest, indicating relatively little power at small spatial scales. The UNM retains more intermediate- and small-scale power, while the CNM has the shallowest spectra over much of the resolved range. The total emission generally follows the CNM more closely than the warmer phases, particularly under compressive forcing. This ordering is consistent with the real-space maps, where the CNM contains the narrowest and strongest structures.

Turbulent forcing further changes the distribution of power across scales. Compressive forcing produces the largest total and CNM amplitudes and the strongest excess of intermediate- and small-scale power. The mixed and solenoidal cases show a smoother distribution of power across scales, with larger relative contributions from the WNM and UNM. In the solenoidal case, the contrast between the relatively shallow CNM spectrum and the steep WNM spectrum remains clear. The similar behavior of $EE$ and $BB$ indicates that the phase- and forcing-dependent redistribution of structure affects both polarization modes.

\subsection{One-point and scale-dependent intermittency}
\label{sec:pdfresults}
\begin{figure*}
\centering
\includegraphics[width=0.95\linewidth]{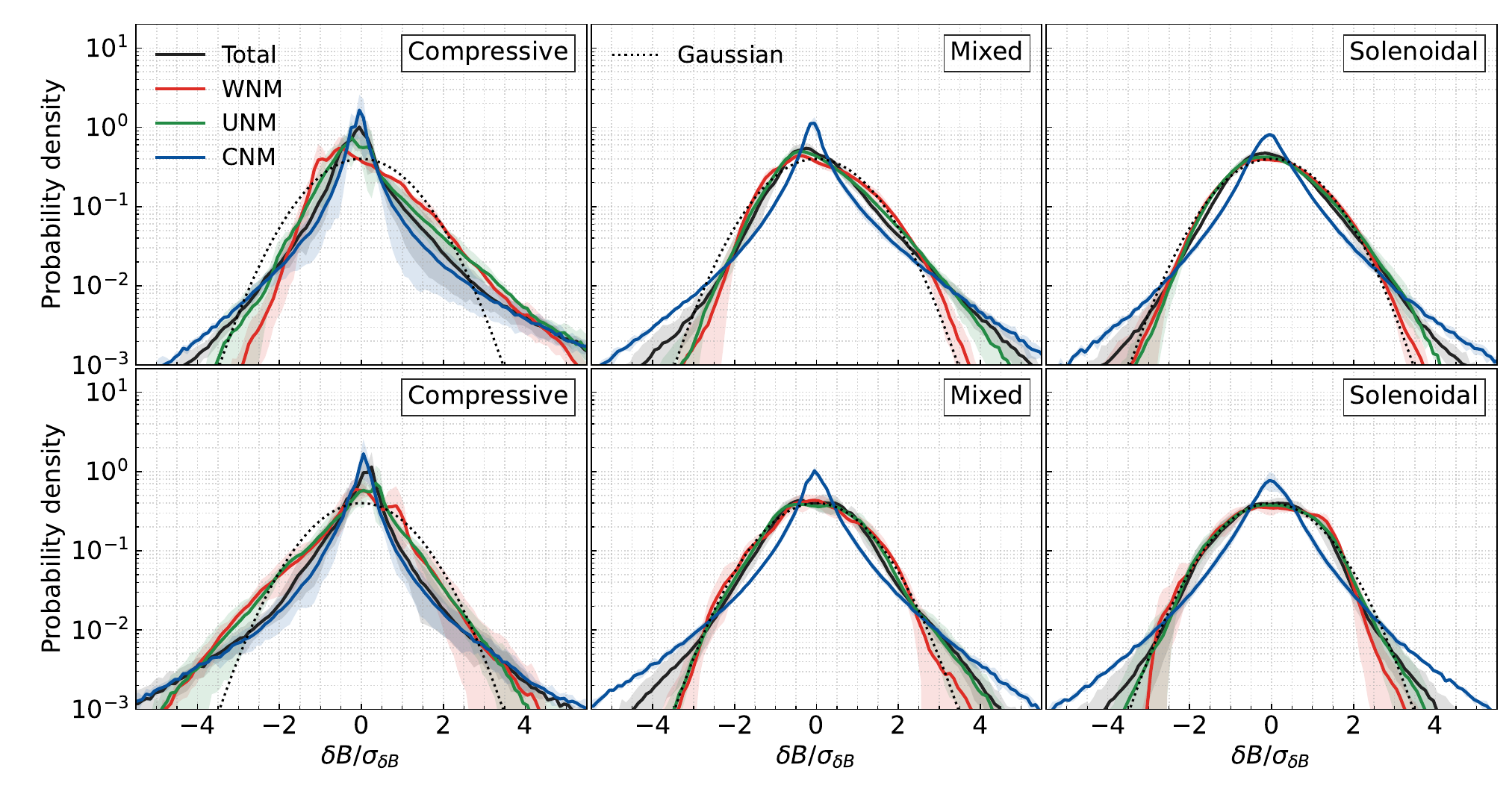}
\caption{One-point PDFs of standardized dust \(E\) modes (top) and \(B\) modes (bottom). Columns show compressive, mixed, and solenoidal forcing. Black, red, green, and blue denote total, WNM, UNM, and CNM emission; the dotted curve is a unit Gaussian.}
\label{fig:onepoint}
\end{figure*}

Fig.~\ref{fig:onepoint} shows that both $E$- and $B$-mode fluctuations are non-Gaussian and clearly depend on thermal phase. The WNM is generally closest to a Gaussian distribution, the UNM shows broader tails, and the CNM has the strongest departures from Gaussianity. The total emission often reflects the CNM contribution. Differences among forcing cases are also present, but their ordering is not uniform across phases and the two polarization modes.

\begin{figure*}
\centering
\includegraphics[width=0.95\linewidth]{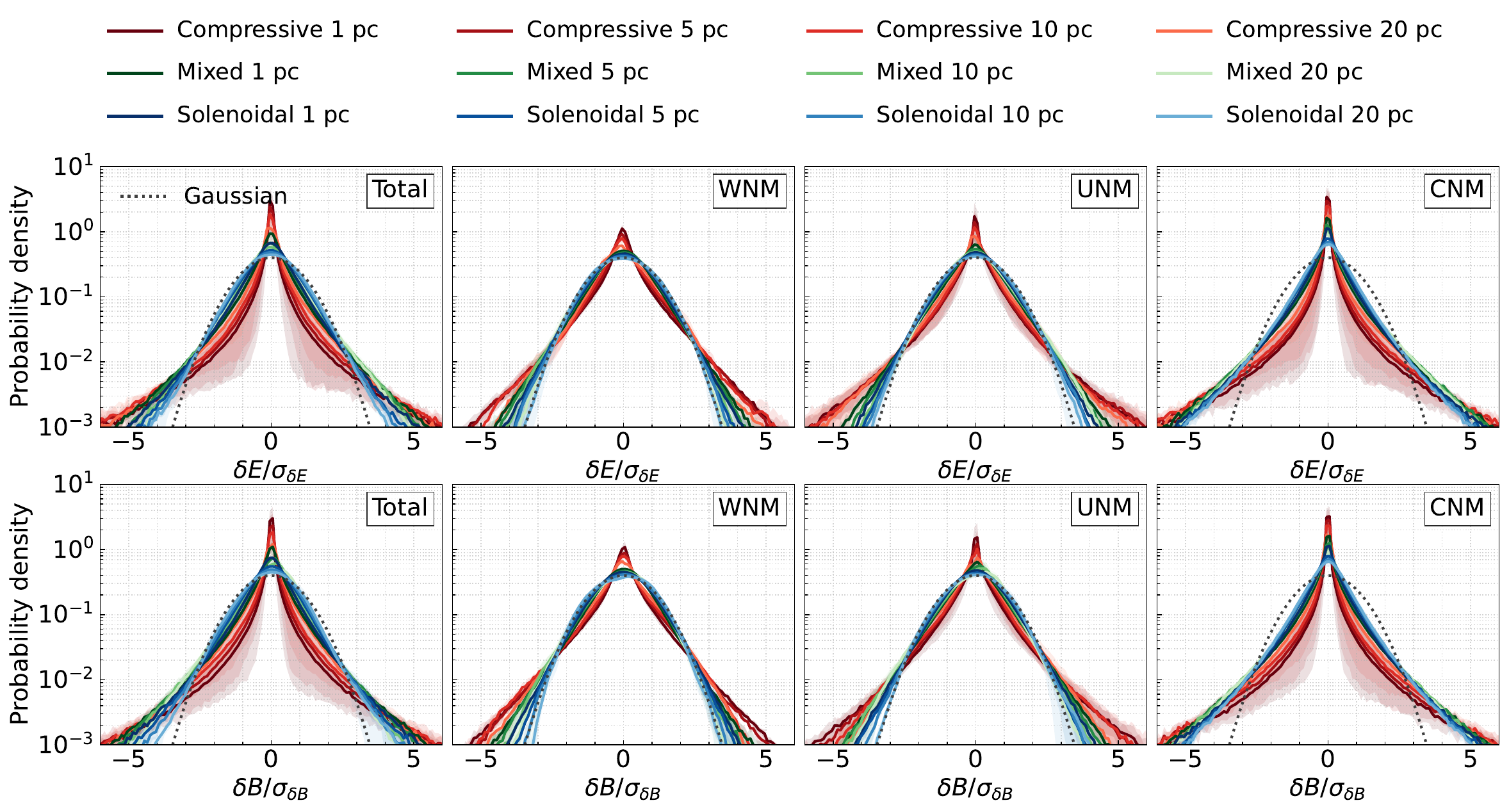}
\caption{PDFs of standardized \(E\)-mode increments (top) and \(B\)-mode increments (bottom) at lags 1, 5, 10, and 20 pc. Columns show total, WNM, UNM, and CNM emission. Red, green, and blue denote compressive, mixed, and solenoidal forcing; line saturation or style identifies lag, and the dotted reference is Gaussian. The two in-plane increment directions are standardized separately and averaged within each snapshot.}
\label{fig:increments}
\end{figure*}

Fig.~\ref{fig:increments} shows a clearer forcing dependence at small spatial scales. At $\ell\simeq1$ pc, compressive forcing generally produces the strongest non-Gaussian increment PDFs, particularly for the total emission and CNM, while mixed and solenoidal forcing show weaker departures. Across all forcing cases, the WNM has the weakest scale dependence and the CNM the strongest. As the lag increases toward $20$ pc, the increment PDFs become progressively closer to Gaussian. The same qualitative behavior is present in both $E$ and $B$.

\subsection{Spatial morphology of intermittency}
\label{sec:morphology}
\begin{figure*}
\centering
\includegraphics[width=0.99\linewidth]{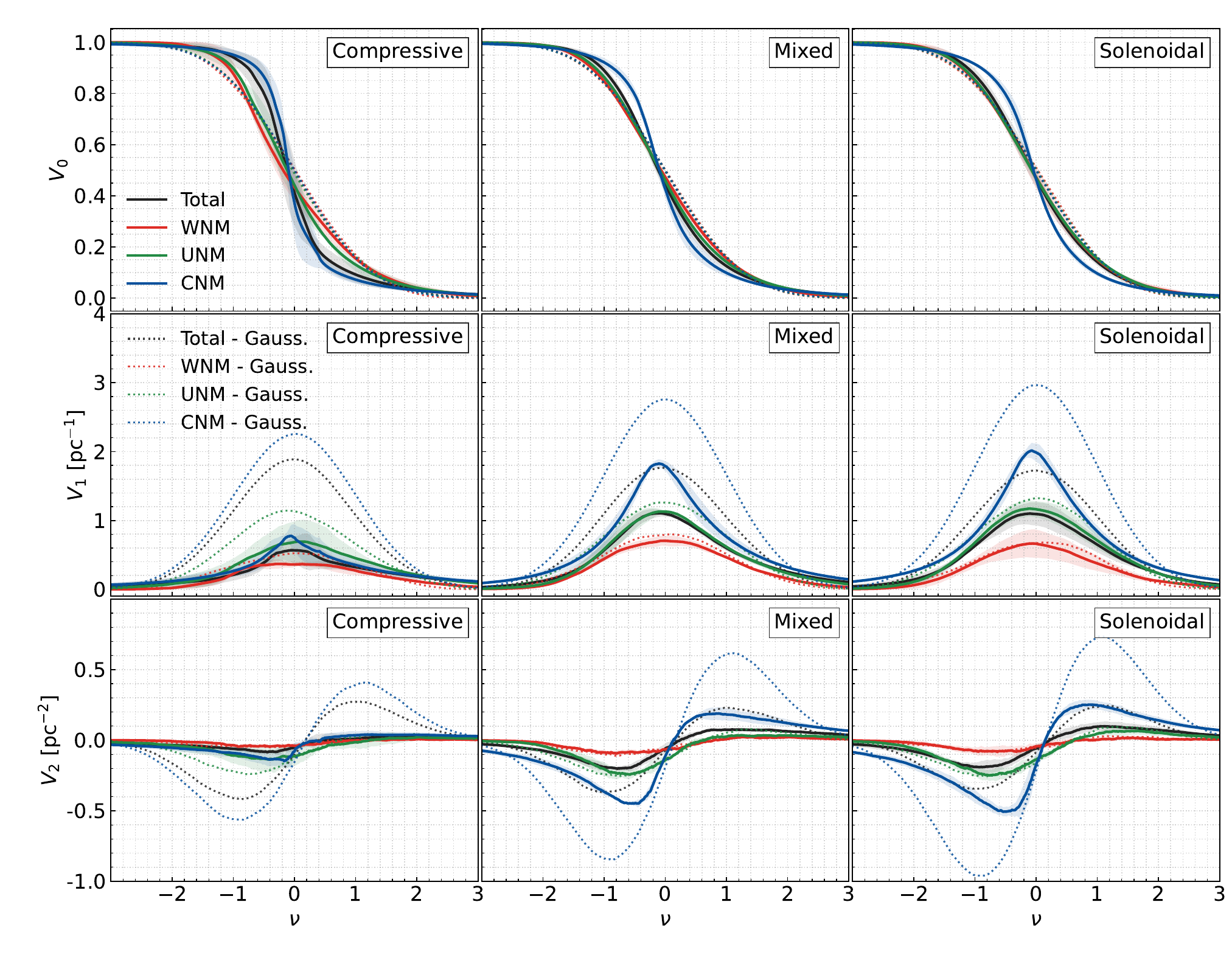}
\caption{Minkowski functionals of standardized dust \(E\)-mode maps as functions of threshold \(\nu\). Rows show excursion-set area \(V_0\), contour length \(V_1\), and Euler characteristic \(V_2\); columns show compressive, mixed, and solenoidal forcing. Black, red, green, and blue denote total, WNM, UNM, and CNM emission. Solid curves are MHD maps and dotted curves are random-phase controls with identical Fourier amplitudes. }
\label{fig:minke}
\end{figure*}

\begin{figure*}
\centering
\includegraphics[width=0.99\linewidth]{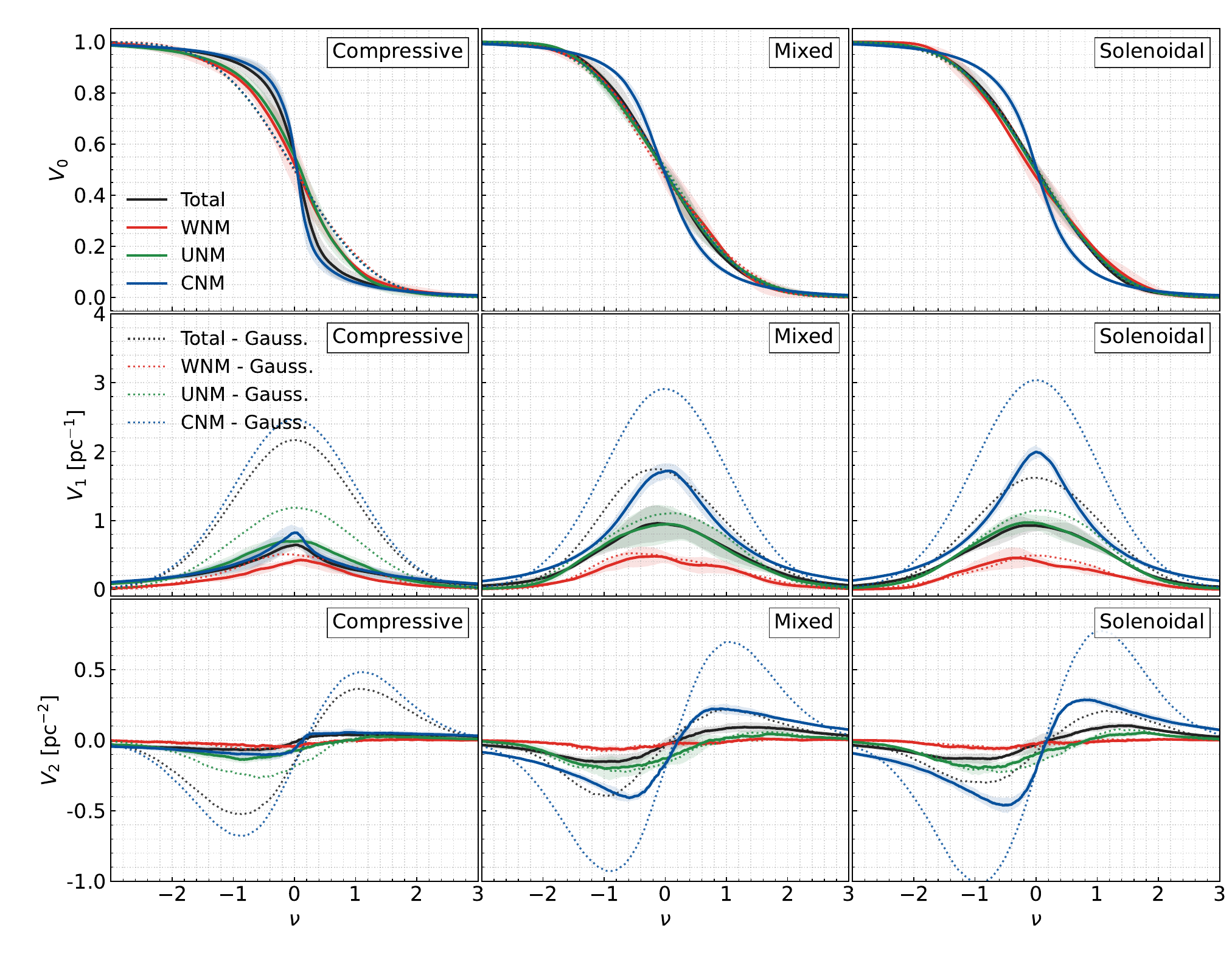}
\caption{Same as Fig.~\ref{fig:minke}, but for \(B\)-mode maps.}
\label{fig:minkb}
\end{figure*}

Figs.~\ref{fig:minke} and \ref{fig:minkb} compare the three Minkowski functionals of the simulated $E$- and $B$-mode maps with random-phase Gaussian reference fields that preserve the same Fourier amplitudes, and therefore the same power spectra.

$V_0(\nu)$ gives the area fraction of the map above a threshold $\nu$ and is closely related to the one-point distribution. The WNM remains closest to the Gaussian reference, whereas larger departures appear in the UNM, CNM, and total emission. The CNM shows the clearest differences at large positive and negative thresholds, indicating that extreme $E$- and $B$-mode fluctuations occupy a different fraction of the map than in Gaussian fields with the same power spectrum. The forcing dependence is visible but varies with phase and threshold, and is less systematic than the thermal-phase dependence.

$V_1(\nu)$ measures the total contour length separating regions above and below the threshold. All phases show some reduction relative to their Gaussian references, but the departure is much stronger in the CNM and total emission than in the WNM. The smaller $V_1$ implies less total boundary length at a given threshold, consistent with the extreme $E$- and $B$-mode fluctuations being organized into larger or more spatially coherent regions than in the random-phase fields. The reduction is particularly strong for compressive forcing in the total and CNM maps, while the mixed and solenoidal cases also show clear departures.

$V_2(\nu)$ measures the topology of the excursion regions through the Euler characteristic. At positive thresholds, it is sensitive to isolated positive $E$ or $B$ regions, whereas at negative thresholds it reflects the topology associated with strong negative fluctuations. The WNM remains close to the Gaussian reference, the UNM shows moderate departures, and the CNM exhibits the largest changes in the amplitude and shape of $V_2$. The total emission shows a similar but weaker behavior. Unlike $V_1$, the forcing dependence of $V_2$ is not monotonic: the relative departures of the compressive, mixed, and solenoidal cases vary with threshold and phase.

Taken together, the Minkowski functionals show a thermal-phase dependence: the WNM is closest to Gaussian morphology, while the CNM carries the strongest higher-order spatial structure. The forcing dependence is present but statistic-dependent. These results show that the non-Gaussianity of the polarized emission involves not only its amplitude distribution, but also the spatial coherence and topology of the $E$- and $B$-mode fluctuations.

\begin{figure*}
\centering
\includegraphics[width=0.95\linewidth]{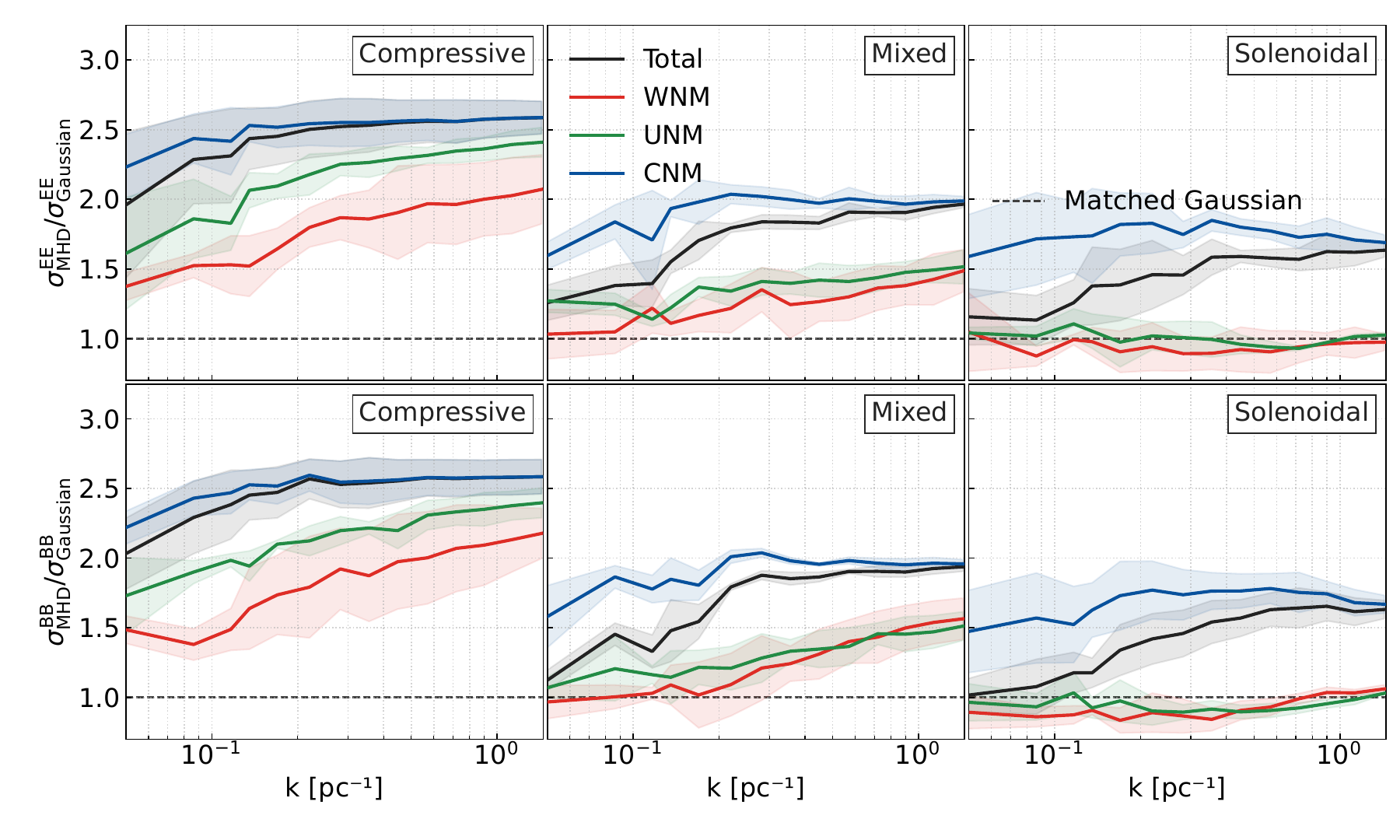}
\caption{Enhancement of spatial dust-power fluctuations relative to a covariance-matched Gaussian foreground. Columns show compressive, mixed, and solenoidal forcing; rows show \(EE\) and \(BB\). Black, red, green, and blue denote total, WNM, UNM, and CNM emission. The ordinate is the ratio of the standard deviations of 25-pc subregion band powers; unity marks Gaussian agreement.}
\label{fig:kvariance}
\end{figure*}

\begin{figure*}
\centering
\includegraphics[width=0.99\linewidth]{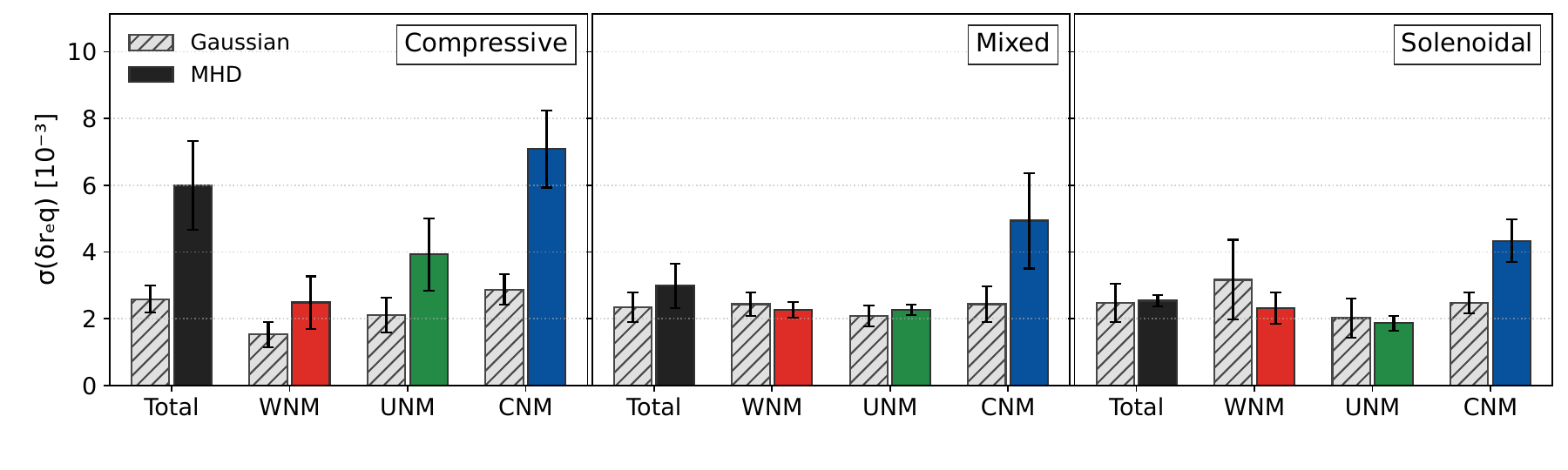}
\caption{Spatial scatter of equivalent tensor contamination for the three forcing mechanisms. Each panel groups total, WNM, UNM, and CNM emission. Solid bars show MHD maps and hatched bars show covariance-matched Gaussian controls. The parent map is divided into 25-pc subregions; physical wavenumbers are mapped to \(60\leq\ell<360\) at 286.48 pc. Error bars show snapshot variation.}
\label{fig:rgeometry}
\end{figure*}
\subsection{Gaussian foregrounds underestimate field-to-field variation}
\label{sec:variance-results}
Fig.~\ref{fig:kvariance} shows how non-Gaussian structure affects the field-to-field variation of polarized dust power. The ordinate is the ratio of the standard deviation of subregion band powers in the MHD maps to that in Gaussian fields with the same two-point covariance, $\sigma_{\rm MHD}^{XX}/\sigma_{\rm Gaussian}^{XX}$, with $X\in\{E,B\}$. A ratio of unity corresponds to the covariance-matched Gaussian expectation, while values above unity indicate additional spatial variation associated with higher-order structure.

The enhancement depends on both thermal phase and turbulent forcing. Under compressive forcing, the total and CNM emission show the largest excess variance, with the ratio reaching $\sim2$--$2.6$ across the resolved wavenumbers. The WNM and UNM also exceed the Gaussian expectation, but by smaller factors. For mixed forcing, the enhancement remains strongest in the CNM, while the total and warmer phases show weaker excess variance. Under solenoidal forcing, the CNM still shows a clear enhancement, whereas the WNM
and UNM remain close to the Gaussian expectation. Similar trends are present in both $EE$ and $BB$. The increase toward higher wavenumbers in the compressive case is consistent with the heavy small-lag increment tails in Fig.~\ref{fig:increments}, linking small-scale intermittency to enhanced field-to-field power fluctuations.

Fig.~\ref{fig:rgeometry} expresses the same effect as an equivalent primordial tensor contamination. For compressive forcing,
$\sigma(\delta r_{\rm eq})$ is larger than in the Gaussian control by factors of $2.31\pm0.26$ for the total emission and $2.47\pm0.11$ for the CNM. The corresponding CNM enhancement remains $2.01\pm0.14$ for mixed forcing and $1.76\pm0.22$ for solenoidal forcing. By contrast, the WNM and UNM are much closer to the Gaussian expectation in the mixed and solenoidal cases, with ratios near or below unity within the snapshot-to-snapshot scatter.

The physical-wavenumber and equivalent-$r$ analyses therefore give the same overall picture. The CNM carries the strongest non-Gaussian spatial variance for all three forcing types, while compressive forcing allows these cold, intermittent structures to dominate the total polarized emission. A Gaussian foreground model can therefore reproduce the mean power and two-point
covariance while still underestimating the field-to-field variation relevant to CMB foreground measurements.

\section{Discussion}
\label{sec:discussion}

\subsection{From three-dimensional intermittency to polarized dust emission}
\label{sec:paper1link}
\cite{Hu2026PaperI} showed that intermittency in the multiphase ISM depends on both thermal phase and turbulent forcing, but that these dependences differ among density, velocity, and magnetic fluctuations. In particular, magnetic intermittency shows the clearest forcing dependence, and the strongest phase-conditioned magnetic intermittency occurs in the UNM rather than the CNM. Extreme magnetic increments preferentially occur where magnetic curl and magnetic tension are simultaneously enhanced, indicating that strong magnetic gradients and field-line distortion coexist.

The projected dust polarization exhibits a different phase ordering. Both $E$ and $B$ are intermittent: their one-point distributions develop non-Gaussian tails, their increment PDFs become increasingly non-Gaussian toward small scales, and their Minkowski functionals depart from random-phase Gaussian fields with the same power spectra. These signatures are weakest in the WNM and strongest in the CNM. In particular, the CNM shows the broadest increment tails (Fig.~\ref{fig:increments}) and the largest changes in contour length and topology (Figs.~\ref{fig:minke} and \ref{fig:minkb}), indicating that its extreme $E/B$ fluctuations are not only stronger but also more spatially organized than expected from the power spectrum alone. The same CNM-dominated hierarchy appears in the excess field-to-field power variance (Fig.~\ref{fig:kvariance}). Compressive forcing generally strengthens the projected intermittency, particularly in the CNM and total emission, although the detailed forcing dependence varies among the different statistics.

This change in phase ordering does not imply that the CNM has the strongest intrinsic magnetic intermittency. Dust polarization weights the underlying MHD fluctuations by gas density and projected magnetic-field geometry before integrating along the line of sight. Dense CNM structures can therefore make a disproportionately large contribution to the observable $E$- and $B$-mode fluctuations. This interpretation is consistent with Paper~I \citep{Hu2026PaperI}, which found that compressive driving increases the CNM mass fraction and redistributes density and magnetic power preferentially into cold gas. The UNM and CNM results are therefore complementary: the UNM hosts the strongest phase-normalized magnetic intermittency, whereas the CNM dominates the strongest emission-weighted and spatially organized polarization fluctuations.

\subsection{Interpretation of the equivalent-\texorpdfstring{$r$}{r} statistic}
\label{sec:rmeaning}

For compressive forcing, the scatter in the equivalent tensor contamination,
$\sigma(\delta r_{\rm eq})$, exceeds that of the covariance-matched Gaussian
control by factors of $2.31\pm0.26$ for the total emission and
$2.47\pm0.11$ for the CNM. Thus, matching the mean dust power and the full
two-point covariance does not reproduce the field-to-field distribution of
foreground contamination when higher-order structure is present.

This result builds on previous evidence that polarized Galactic dust contains
substantial non-Gaussian information beyond its power spectrum.
Wavelet-scattering analyses of synthetic and observed dust polarization have
identified multiscale phase correlations and filamentary structure that are
not captured by two-point statistics
\citep{RegaldoSaintBlancard2020,Delouis2022}. The present analysis connects
this higher-order structure to the thermal phases and turbulent forcing of the
underlying ISM, and shows that it also broadens the spatial distribution of
measured $E$- and $B$-mode power.

The particularly strong enhancement in the compressive case is consistent with the three-dimensional behavior identified in Paper~I \citep{Hu2026PaperI}. Compressive driving increases the CNM mass fraction and redistributes density and magnetic power preferentially into cold gas. After density weighting and projection, these concentrated CNM structures make a large contribution to the spatial variation of the polarized foreground. The equivalent-$r$ statistic therefore measures an observational consequence of this redistribution rather than the intrinsic intermittency of the CNM alone.

The quoted factors are not direct uncertainties on a measurement of $r$. The final uncertainty of a CMB experiment also depends on sky fraction, instrumental noise, lensing, frequency coverage, masks, transfer functions, and component separation. Here, $\sigma(\delta r_{\rm eq})$ isolates the foreground scatter associated with higher-order dust structure under a controlled mean normalization. Its absolute value depends on the adopted dust normalization, angular mapping, subregion size, and multipole range. The ratio between the MHD and Gaussian cases is less sensitive to a common amplitude rescaling and therefore provides a more direct measure of the additional variance associated with non-Gaussianity. In this sense, the factor of $\sim2.5$ for the compressive CNM means that the $r$-weighted foreground distribution among subregions is approximately 2.5 times broader than predicted by a Gaussian field with the
same two-point covariance.

Non-Gaussian dust covariance has also been investigated directly in the context of CMB \(B\)-mode inference \citep{AbrilCabezas2024}. That work modeled non-Gaussianity phenomenologically through large-scale modulation and synthetic filamentary dust structure, rather than deriving it from multiphase ISM turbulence. Our calculation instead starts from MHD simulations and connects the resulting foreground variance to thermal phase and turbulent forcing. The two approaches are therefore complementary: their analysis addresses how a prescribed non-Gaussian dust covariance propagates into \(r\) inference, whereas ours isolates a physical ISM origin for the excess foreground variance before instrumental noise, lensing, and multifrequency component separation are included.

\subsection{Connection to observations}
\label{sec:observations}
The simulations predict that the thermal composition of the diffuse ISM should be reflected in the non-Gaussian statistics of polarized dust emission. WNM-dominated regions should remain relatively close to Gaussian,
whereas a larger CNM contribution should be associated with broader tails, larger morphological departures from Gaussian fields, and greater field-to-field power variance. This prediction concerns the emission-weighted projected signal: it does not imply that the CNM has the strongest intrinsic magnetic intermittency. Paper~I instead found the strongest phase-normalized magnetic intermittency in the UNM, while compressive forcing preferentially redistributes density and magnetic power into the CNM \citep{Hu2026PaperI}.

The observational prediction is therefore not simply the presence of non-Gaussian dust structure, which has already been demonstrated in both simulations and Planck polarization maps \citep{RegaldoSaintBlancard2020,Delouis2022}, but its connection to specific thermal phases and turbulent forcing. Phase-sensitive H\,{\sc i} decomposition provides a direct way to test this connection \citep{KalberlaHaud2018}. High-latitude regions with comparable mean 353-GHz dust power can be grouped by their inferred WNM, UNM, and CNM fractions and compared using one-point and increment PDFs, Minkowski functionals, and the field-to-field scatter of $BB$ power. Gaussian or phase-randomized maps matched to the observed two-point statistics provide the corresponding null model. A systematic increase of non-Gaussian power variance
with CNM fraction at fixed mean dust power would directly test the projected trend found here.

The physical origin of these fluctuations can be explored with complementary tracers. Paper~I found that extreme magnetic increments preferentially occur where magnetic curl and magnetic tension are simultaneously enhanced, while extreme density fluctuations are associated with convergence and thermal gradients primarily in the UNM \citep{Hu2026PaperI}. The Velocity Gradient Technique (VGT) provides an independent estimate of the plane-of-sky magnetic orientation from spectroscopic H\,{\sc i} emission \citep{2018ApJ...853...96L,2018MNRAS.480.1333H,2023MNRAS.524.2379H}. Comparing VGT-inferred field orientations with regions of strong dust-polarization increments could help distinguish fluctuations associated primarily with magnetic geometry from those dominated by density weighting. H\,{\sc i} phase information and molecular shock tracers such as SiO can provide complementary constraints on thermal structure and strongly compressive events \citep{Rybarczyk2023,2024ApJ...967...18Z}.

For CMB applications, these results suggest that mean dust amplitude alone may not fully characterize the foreground variability of an observing field. Thermal-phase information, small-scale polarization statistics, and non-Gaussian morphology may provide additional information on the expected field-to-field scatter. Their practical value for field selection and foreground likelihoods will require calibration against observed 353-GHz polarization maps with realistic noise, masks, and component-separation effects.

\subsection{Limitations}
\label{sec:limitations}
The present study isolates how thermal phase and turbulent forcing shape the non-Gaussian statistics of polarized dust emission in a controlled multiphase ISM. A natural next step is to extend the calculation over a broader turbulence parameter space, including a wider range of sonic and Alfv\'enic Mach numbers, forcing mixtures, and mean magnetic-field strengths. Larger simulation volumes and a wider dynamic range would also allow the intermittency and field-to-field variance to be followed across a broader range of spatial scales.

The connection to CMB observations can be developed further through more realistic synthetic skies. Future calculations should include phase-varying grain properties and multifrequency dust emission, together with instrumental beams, realistic noise, sky masks, and lensing $B$ modes. Embedding the simulated dust structure in a component-separation pipeline would then determine how the additional non-Gaussian foreground variance
propagates into practical constraints on $r$.

Finally, comparison with phase-resolved H\,{\sc i} observations and 353-GHz polarization maps can test whether the predicted dependence on CNM fraction, turbulent forcing, and spatial scale is present in the real sky. Such studies would connect the physical origin of ISM intermittency more directly to foreground modeling and CMB field selection.

\section{Conclusions}
\label{sec:conclusions}

Using three thermally multiphase MHD simulations and three snapshots for each forcing case, we have traced ISM intermittency from the three-dimensional turbulent medium into synthetic dust $E$- and $B$-mode polarization. Our main conclusions are:

\begin{enumerate}

\item Both dust $E$ and $B$ modes are intermittent. Their one-point and increment PDFs show non-Gaussian, scale-dependent tails, while their Minkowski functionals reveal spatial organization that differs from random-phase Gaussian fields with the same power spectra.

\item The projected intermittency depends strongly on thermal phase. The WNM is generally closest to Gaussian, whereas the CNM shows the strongest non-Gaussian tails, morphological departures, and excess field-to-field power variance. Compressive forcing enhances these signatures most clearly in the CNM and total emission. This differs from Paper~I, where the strongest phase-normalized magnetic intermittency occurs in the UNM, because dust polarization preferentially weights dense CNM structures.

\item A Gaussian foreground matched to the two-point covariance can substantially underestimate spatial fluctuations in dust power. For compressive forcing, the equivalent-$r$ contamination scatter is larger than the Gaussian prediction by factors of $2.31\pm0.26$ for the total emission and $2.47\pm0.11$ for the CNM. The CNM remains enhanced under mixed and solenoidal forcing, by factors of $2.01\pm0.14$ and $1.76\pm0.22$, respectively.

\item Thermal-phase composition and turbulent forcing therefore carry foreground information beyond the mean $E/B$ power spectra and two-point covariance. In particular, dense cold structures can control the spatial distribution of polarized dust power and broaden the range of foreground contamination encountered among different observing fields.

\end{enumerate}


\begin{acknowledgments}
The Sherman Fairchild Postdoctoral Fellowship at the California Institute of Technology supports Y.H.. This work used SDSC Expanse CPU and NCSA Delta CPU through ACCESS allocations PHY230032, PHY230033, PHY230091, PHY230105, PHY230178, and PHY240183. National Science Foundation supports the ACCESS program grants 2138259, 2138286, 2138307, 2137603, and 2138296.
\end{acknowledgments}

\software{AthenaK \citep{2024arXiv240916053S}}

\newpage
\bibliography{sample631}{}
\bibliographystyle{aasjournal}

\end{document}